\documentclass[preprint,12pt]{elsarticle}

\usepackage{graphicx}

\usepackage[ruled,lined,algonl,boxed]{algorithm2e}

\usepackage{amssymb}
\usepackage{subfigure}

\usepackage{amsthm}

\newtheorem{thm}{Theorem}

\newdefinition{rmk}{Remark}
\newproof{pf}{Proof}
\newproof{pot}{Proof of Theorem \ref{thm2}}

\journal{}

\usepackage{amsopn}
\DeclareMathOperator*{\argmax}{\operatornamewithlimits{arg\,max}}

\begin{document}

\begin{frontmatter}



\title{Multiple domination models for placement of electric vehicle charging stations in road networks}


\author[label2a]{Padraig Corcoran}
\ead{corcoranp@cardiff.ac.uk}
\author[label2]{Andrei Gagarin\corref{cor1}}
\ead{gagarina@cardiff.ac.uk}
\cortext[cor1]{Corresponding author, phone: +44\,(0)2920 688850}
\address[label2a]{School of Computer Science \& Informatics, Cardiff University, Queens Bldg, 5 The Parade, Cardiff CF24 3AA, UK}
\address[label2]{School of Mathematics, Cardiff University, 21-23 Senghennydd Rd, Cardiff CF24 4AG, UK}

\begin{abstract}
Electric and hybrid vehicles play an increasing role in the road transport networks. Despite their advantages, they have a relatively limited cruising range in comparison to traditional diesel/petrol vehicles, and require significant battery charging time. We propose to model the facility location problem of the placement of charging stations in road networks as a multiple domination problem on reachability graphs. This model takes into consideration natural assumptions such as a threshold for remaining battery load, and provides some minimal choice for a travel direction to recharge the battery. Experimental evaluation and simulations for the proposed facility location model are presented in the case of real road networks corresponding to the cities of Boston and Dublin. 
\end{abstract}

\begin{keyword}
Road networks \sep Electric vehicles \sep Facility location problem \sep $k$-Domination \sep
$\alpha$-Domination \sep Heuristic optimization


\end{keyword}

\end{frontmatter}


\section{Introduction}
\label{Intro}
Due to increasing concerns about the environment, the resulting policies and advances in technology, zero and low emission electric and hybrid vehicles are playing an ever more important role in road transportation. Despite the advantages of electric vehicles, their relatively limited cruising range (in comparison to traditional diesel/petrol vehicles) and significant battery charging time often provide major challenges to their usage.

As a result, in order for electric vehicles to be viable, it is necessary to have a sufficient number of charging stations which are appropriately distributed throughout a road network. Given a particular road network layout, determining appropriate locations and capacities for such charging stations is a challenging multi-objective optimisation problem with many constraints. One of the key objectives is to minimise the length of detours from a desired route which are necessary for recharging. On the other hand, constraints in this optimisation problem include requiring the number of charging stations to be reasonably small, ensuring the distance between consecutively used stations does not exceed the cruising range of electric vehicles, and that the capacities of the charging stations be sufficient enough to avoid bottlenecks. In this article, we focus on the problem of optimising the placement of charging stations such that the length of detours necessary for recharging is minimised subject to the constraint that the number of charging stations is reasonably small.

In the existing literature (e.g., see \cite{FNS2015}), the problem of charging station placement is often modelled as a shortest-path vertex cover problem for graphs. In this model, a vehicle is assumed to begin with a fully charged battery and follow a shortest path from an initial point to a final destination without much deviation. However, in many cases this assumption is not going to be valid, and the model in question is not going to be suitable. For example, mail or groceries delivery drivers are usually concerned with navigating in a way prescribed by delivery options (in time and space), and are not particularly concerned about shortest paths issues when navigating a certain area. Also, traffic jams, road closures and other temporary or sudden obstacles (e.g., a snow storm in Canada) may significantly influence the originally intended shortest path for driving. As a result, it is more natural and plausible to assume that drivers will become concerned about their remaining cruising range and battery charge only after the battery level falls below a certain low threshold, implying the remaining distance they can travel is quite limited.

In this work we propose a novel model for the placement of charging stations in road networks which is based on computing \emph{multiple domination models} for a \emph{reachability graph} corresponding to the original road network. A reachability graph models the set of locations which are reachable from a given location, where a location is reachable if its distance from the location in question is below a certain threshold. The reachability graph appropriately models the situation where a driver becomes concerned about their low battery charge and wishes to make a detour to a recharging station which is reachable from their current location. By considering multiple domination models on the reachability graph, we can compute a set of charging stations locations such that each location in the network can be served from several charging stations. That is, multiple charging stations are reachable from each such location. The driver therefore has several charging stations options to select from and can in turn select the one which minimises the necessary detour.

The layout of this paper is as follows. In Section \ref{sec:related_work} we provide a concise overview of related work. The proposed facility location problem model is presented in Section \ref{sec:model}. Section \ref{sec:algorithm} describes the algorithms used to compute the reachability graph and multiple domination models for the road network, provides some analysis and explains heuristic adjustments for the algorithms. An experimental evaluation using real road networks corresponding to the cities of Boston and Dublin is presented in Section \ref{sec:experimental_evaluation}. Finally, in Section \ref{sec:conclusions} we draw some conclusions and discuss possible future research directions.


\section{Related work}
\label{sec:related_work}
There exist quite a large volume of recent literature related to electric vehicles and optimization in road networks focusing on different aspects of problem modelling and corresponding solution methods. For example, Poghosyan et al. \cite{PGHL2015} discuss possible scenarios of distribution of loads in the power grids and their dependence on temporal, spatial, and behavioural charging patterns for electric vehicles.

Given a set of charging stations and their locations fixed in the network, the authors in \cite{SF2013} propose a method for computing all locations which are reachable from a given initial location, assuming a specified number of battery recharges can be done. In this work, the locations of the charging stations are assumed to be fixed, and there is no attempt to optimize the placement of charging stations in the network.

In \cite{LLC2014}, the authors consider a specific type of the general facility location problems called the electric vehicle charging station placement problem. In their work, they try to minimize construction costs for placement of charging stations in few pre-selected locations subject to a set of  constraints. The problem is modelled by using mixed-integer linear programming (MILP) with some non-linear constraints. The authors show that the problem is NP-hard and propose several solution methods by reduction to MILP problems and using heuristics. An experimental evaluation is first done with randomly generated small-size synthetic instances using MATLAB and generic MILP solvers. Then the model and methods are evaluated in the case of possible scenarios of building charging stations in Hong Kong by considering $18$ pre-selected locations for potential construction of charging stations corresponding to different districts of the country.
Notice that, in this model, the sites for potential construction of charging stations are pre-selected, and the average cruising distance of fully-charged electric vehicles is used to select the sites minimizing the total construction costs. 

In \cite{FNS2015}, the authors model the problem of placement of charging stations as a ``shortest path" cover problem in a graph of the road network $G=(V,A)$. One needs to find a smallest subset of vertices $L\subseteq V$ such that every minimal shortest path in $G$ that exceeds the electric vehicle battery capacity has a loading station placed in a vertex of the set $L$.
The problem is then modelled as a special type of the Hitting Set problem: the collection of subsets of $V$ to be hit by the charging stations corresponds to the minimal shortest paths in $G$ that exceed the battery capacity.
An adaptation of the standard greedy approach provides an $O(\log|V|)$-approximation algorithm to solve this problem.
The instance construction and representation are described as the main challenges with respect to using limited computational memory and time resources. As a result, using different representations and searching for minimal shortest paths turns out to be a quite complicated task and is involved with many  details. Overall, the problem does not seem to scale well, and the heuristic improvements for the implementation would be very challenging to reproduce.


\section{Reachability graph and multiple domination models}
\label{sec:model}
\begin{figure}
\centering     
\subfigure[]{\label{fig:b}\includegraphics[width=6.4cm]{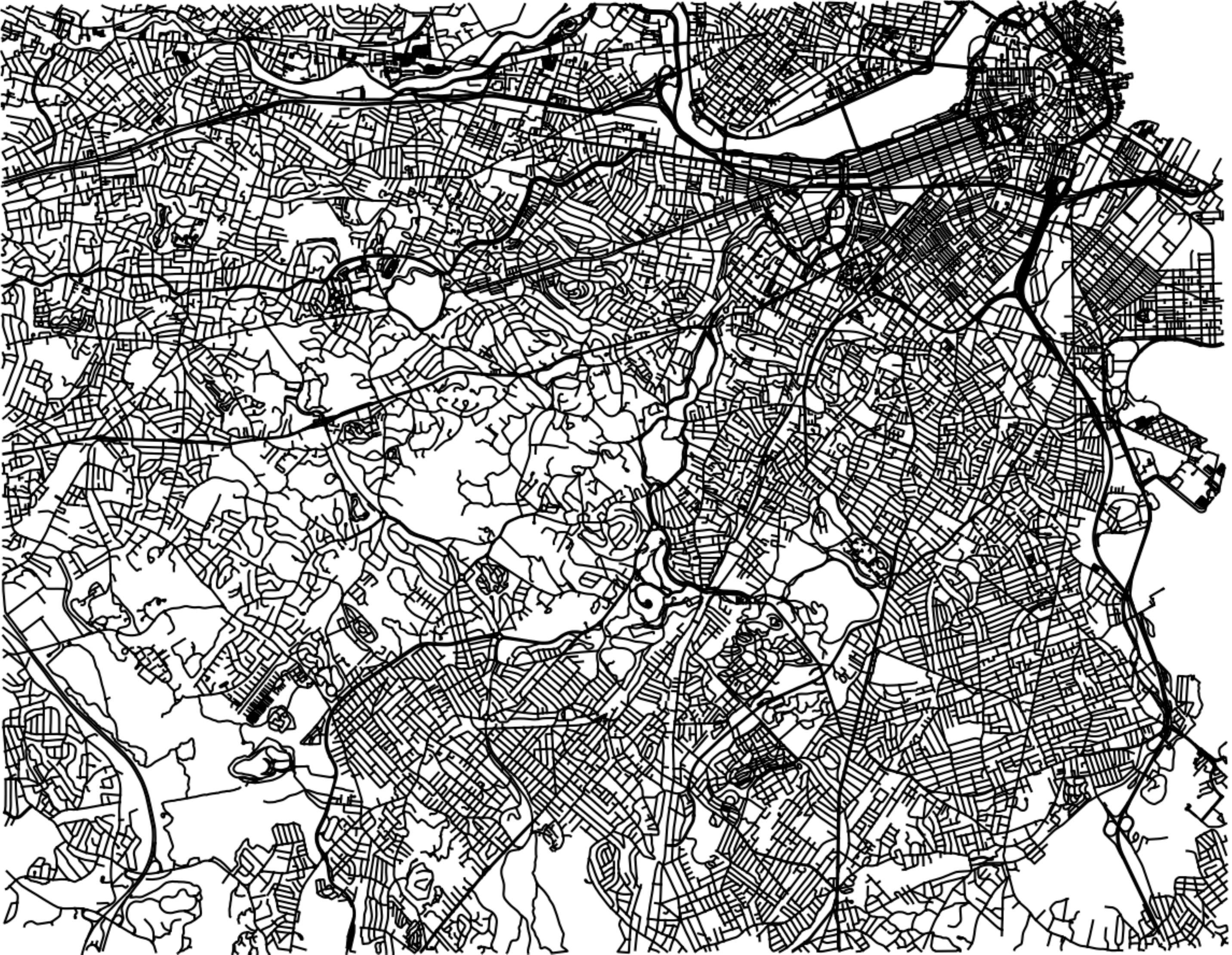}
\label{fig:Boston_street_network}}
\hspace{1mm}
\subfigure[]{\label{fig:a}\includegraphics[width=6.4cm]{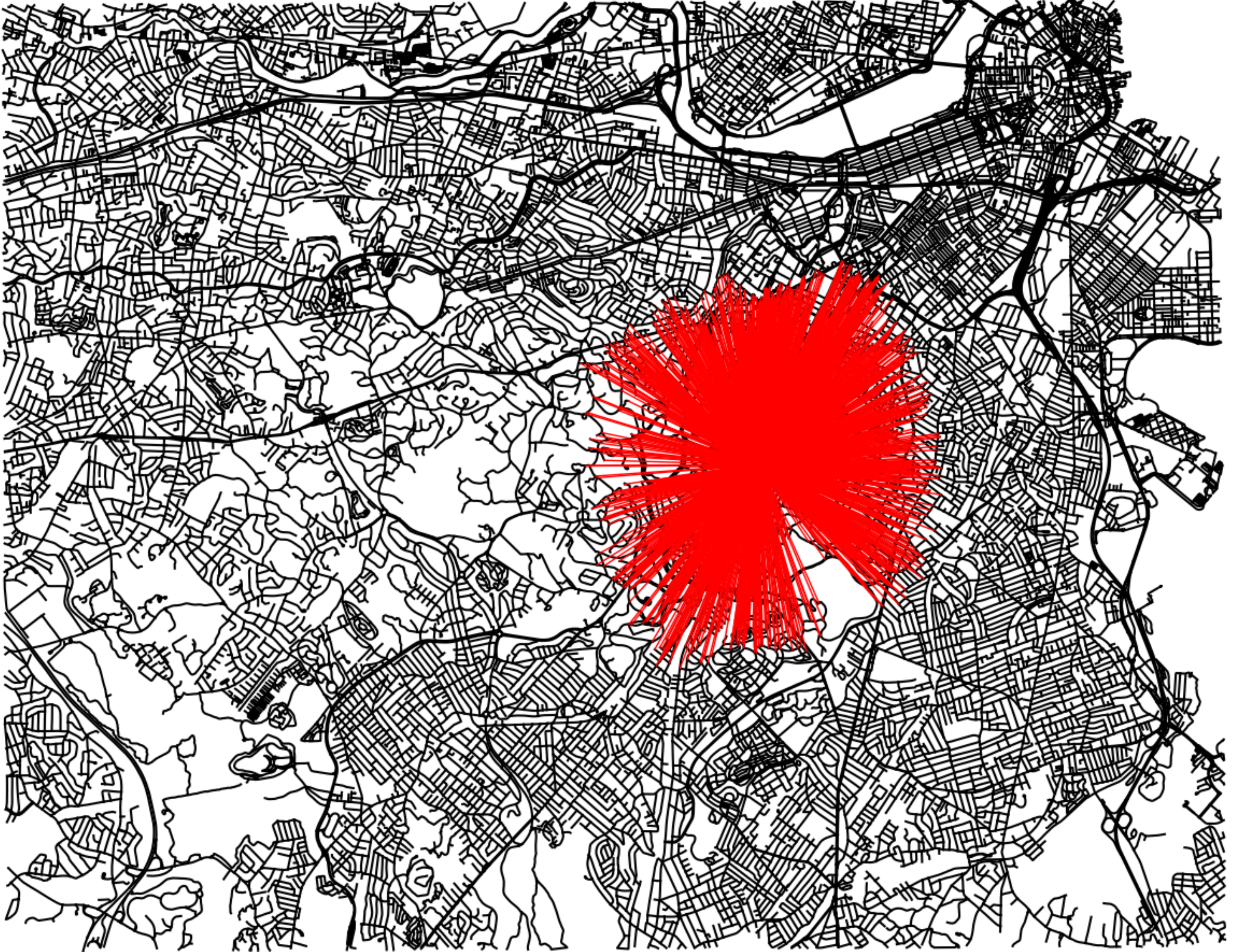}
\label{fig:reachability_vertex}}
\caption{(a)\,The road network for the city of Boston; (b)\,Neighbourhood of a vertex in the corresponding reachability graph.}
\label{fig:Boston_reachability_vertex}
\end{figure}

For simplicity, we consider a road network represented by a weighed undirected simple graph $G^s=(V^s,E^s, w:E^s\rightarrow \mathbb{R})$, where the set of vertices $V^s$ corresponds to road intersections and dead-ends, while the set of edges $E^s$ corresponds to road segments connecting these vertices. The weight $w(e)$ on an edge $e\in E^s$ is the length of the corresponding road segment (in meters). An example of this graph model for the road network of the city of Boston is illustrated in Figure \ref{fig:Boston_street_network}.

Given a road network graph $G^s=(V^s,E^s, w:E^s\rightarrow \mathbb{R})$, we define its \emph{reachability graph} $G^r_t=(V^r,E^r_t)$ as a simple (unweighted) graph with $V^r = V^s$ and edges $uv\in E^r_t$ if and only if the length of shortest path (distance) between the corresponding vertices $u$ and $v$ in $G^s$ is less than a specified \emph{reachability threshold} of $t$ km, i.e. $w(P^s_{uv})\le t$, where $P^s_{uv}$ and $w(P^s_{uv})$ are a shortest path and corresponding distance between $u$ and $v$ in $G^s$, respectively. The reachability graph corresponding to the Boston road network of Figure \ref{fig:Boston_street_network} for $t=3.0$km is illustrated in Figure \ref{fig:reachability_vertex}. In this figure, red line segments are drawn between a given vertex and each of its neighboring vertices in $G^r_{3.0}$. The reachability graph $G^r_t$ appropriately models the situation where a driver becomes concerned about their low battery and wishes to make a detour to a recharging station which is reachable from their current location.

Notice that the reachability threshold $t$ to construct a reachability graph $G^r_t$ should normally satisfy the following lower bound derived from the road network graph $G^s$:
\begin{equation}
\label{bound:max_min}
t\ge \max_{u\in V^s} \min_{v\in N(u)} {w(uv)},
\end{equation}
where $N(u)$ is a set of all vertices adjacent to $u$ in $G^s$, and $uv\in E^s$.
In other words, from any given point $u\in V(G^s)$, it should be possible to reach at least one of the neighbouring locations $N(u)$ using the remaining battery power (to eventually recharge the battery).
This would imply the reachability graph $G^r_t$ has no isolates.
Similarly, for better flexibility, more choice, and ``safer" conditions for reaching possible recharging locations, one may impose the stronger lower bound for the threshold
\begin{equation}
\label{bound:max}
t\ge \max_{e\in E^s} {w(e)}.
\end{equation}
This would mean it is possible to reach all the neighbouring locations $N(u)$ from any given point $u\in V(G^s)$ using the remaining battery power. The lower bound (\ref{bound:max}) would imply the vertex degrees of $G^r_t$ are at least the corresponding vertex degrees of $G^s$.

However, in the case of a small number of remote locations which are more difficult to reach in the network, it may be too demanding and expensive to satisfy the lower bound (\ref{bound:max}) or even  (\ref{bound:max_min}) for the whole network. Therefore, when conditions of the lower bound (\ref{bound:max}) or (\ref{bound:max_min}) are not satisfied, the remote locations (``outliers" of the road network) should be treated separately. Thus, the ``outliers" are considered in our models as well.

Having constructed a road network graph $G^s$ and a corresponding reachability graph $G^r_t$, the problem of placing charging stations in the road network becomes a facility location problem which can be modelled on the graphs $G^s$ and $G^r_t$ as follows.
In general, if $G$ is a graph of order $n$, then
$V(G)=\{v_1,v_2,...,v_n\}$ is the set of vertices of $G$, the degree of vertex $v_i$ is denoted by $d_i$ or $d(v_i)$, $i=1,\ldots,n$, 
the minimum and maximum vertex degrees of $G$ are denoted by $\delta=\delta(G)$ and $\Delta=\Delta(G)$, respectively. 
The neighbourhood of a vertex $v$ in $G$ is denoted by $N(v)$. 
A subset $X\subseteq V(G)$ is
called a {\it dominating set} of $G$ if every vertex not in $X$ is
adjacent to at least one vertex in $X$. The minimum cardinality of
a dominating set of $G$ is called the {\it domination number} of $G$ and denoted by 
$\gamma(G)$.
Dominating sets in graphs are natural general models for facility location problems in networks.

Given an integer $k\ge 1$, a set $X\subseteq V(G)$ is called a {\it $k$-dominating set} of $G$ if
every vertex $v\in V(G)\backslash X$ has at least $k$ neighbours in $X$. The
minimum cardinality of a $k$-dominating set of $G$ is the {\it
$k$-domination number} $\gamma_k(G)$. Clearly, $\gamma_1(G)=\gamma(G)$, and $\gamma_{k_1}(G)\le\gamma_{k_2}(G)$ when $k_1\le k_2$.
Given a real number $\alpha$, $0<\alpha\le1$, a set
$X\subseteq V(G)$ is called an {\it $\alpha$-dominating set} of
$G$ if for every vertex $v\in V(G)\backslash X$, $|N(v)\cap X|\ge \alpha
d_v$, i.e. $v$ has at least $\lceil\alpha d_v\rceil$ (i.e. $\alpha\times 100\%$)
neighbours in $X$. The minimum cardinality of an $\alpha$-dominating
set of $G$ is called the {\it $\alpha$-domination number}
$\gamma_{\alpha}(G)$. 
It is easy to see that
$\gamma(G)\le\gamma_{\alpha}(G)$, and
$\gamma_{\alpha_1}(G)\le\gamma_{\alpha_2}(G)$ for
${\alpha_1}<{\alpha_2}$. Also, $\gamma(G)=\gamma_{\alpha}(G)$ when
$\alpha$ is sufficiently close to $0$.

The $k$- and $\alpha$-domination are two types of multiple domination in graphs. The concept of $\alpha$-domination differs from the $k$-domination in that a vertex must be dominated by a certain percentage ($\alpha\times 100\%$) of the vertices in its neighbourhood instead of a fixed number $k$ of its neighbours.
Each of these two types of multiple domination can be used to model the situation when an electric vehicle driver starts to look for a conveniently located battery charging station and needs to have several options where to recharge the battery.
In this paper, we focus on $k$-domination, which means that in any location (vertex) of the network (graph) the driver can use one out of $k$ possible options, $k=1,2,\ldots,\delta$. Clearly, in the case $k>\delta$, this model suggests that the vertices of degree less than $k$ are all included into the $k$-dominating set or ignored (i.e. treated separately). Therefore, without loss of generality, we can assume $k\le\delta$. 

The problems of finding exact values of $\gamma_k(G)$ and $\gamma_{\alpha}(G)$ are known to be NP-complete \cite{LC2013,DHLM2000}. Therefore, it is important to have efficient heuristic algorithms and methods to find small-size $k$- and $\alpha$-dominating sets in graphs. Also, it is important to have good theoretical bounds for $\gamma_k(G)$ and $\gamma_{\alpha}(G)$ to be able to estimate quality of a given solution set.
The following two general upper bounds for the $k$- and $\alpha$-domination numbers have been obtained in \cite{GPZ2009,GPZ2013} by using a probabilistic method approach. These bounds generalize a classic upper bound for the domination number $\gamma(G)$. Also, the probabilistic constructions used in the proofs of these bounds allow us to design randomized algorithms to find $k$- and $\alpha$-dominating sets such that the expected order of the set of vertices returned by the algorithm satisfies the corresponding upper bound.

Putting
$
\delta' = \delta -k +1
$
and 
$
 {b}_{k-1} = \pmatrix{\delta \cr k-1},
$
where $0\le k-1\le \delta$, we have:
\begin{thm}[\cite{GPZ2013}]\label{k-dom}
 For every graph $G$ with $\delta\ge k$,
$$
\gamma_{k}(G) \le \left(1-{\delta' \over {b}_{k-1}^{\ 1/ \delta'}
{(1+\delta')}^{1+1/\delta'}}\right) n.
$$
\end{thm}
For $0<\alpha\le 1$, we put 
$\widehat{\delta} = \lfloor \delta (1 - \alpha)
\rfloor +1$ and  
$\displaystyle{
{\widehat d}_{\alpha} = {1\over n}
\sum_{i=1}^n \pmatrix{d_i \cr {\lceil \alpha d_i \rceil -1}}}.
$
Then we have:
\begin{thm}[\cite{GPZ2009}]\label{main}
For every graph $G$,
$$
\gamma_{\alpha}(G) \le \left(1-{\widehat{\delta} \over
{{\widehat d}_{\alpha}^{\;1/\widehat{\delta}}(1+\widehat{\delta})}^{1+1/\widehat{\delta}} } \right) n.
$$
\end{thm}

Clearly, given a reachability graph $G^r_t$, increasing the reachability threshold $t$ can only extend the neighbourhoods of vertices in $G^r_t$ to obtain $G^r_q$, $q>t$, i.e. $G^r_t$ is a spanning subgraph of $G^r_q$.
Therefore, given a $k$-dominating set $X\subseteq V(G^r_t)$ in $G^r_t$, one can infer some properties about this set $X\subseteq V(G^r_q)$ in the reachability graph $G^r_q$, where $q \geq t$. 
Clearly, having $k$ and the set $X$ fixed, $k\le |X|$, and every vertex $v\in V(G^r_q)\backslash X$ is dominated by at least $k$ vertices in $X$.
Then, as the reachability threshold $q$ increases, keeping the set $X$ fixed and considering $k$ as a parameter, the number $k$ can be eventually increased.
When the reachability threshold $q$ is at least the diameter of the network graph $G^s$, the reachability graph $G^r_q$ becomes a complete graph, and every vertex $v\in V(G^r_q)\backslash X$ is dominated by all the vertices in $X$, so that we can set $k=|X|$. 

If $q \leq t$, one cannot infer any domination properties of the $k$-dominating set $X$ of $G^r_t$ in the (spanning) reachability graph $G^r_q$. However, as $q$ approaches $t$, the set $X$ is going to start to behave like a $k$-dominating set with respect to the reachability graph $G^r_q$, and eventually the reachability threshold $t$ can be lowered. Some of the above properties are illustrated in the experimental results section of this paper.

\section{Basic algorithms, heuristics, their implementation and complexity analysis}
\label{sec:algorithm}

In this section, we describe the basic algorithmic ideas and routines to compute the reachability graphs and to find $k$-dominating sets in the reachability graphs. They are developed from and based on the theoretical results described in \cite{AS1992,GPZ2013} and our simulations and experiments with real road networks of Dublin and Boston. The $k$-dominating sets in the reachability graphs are facility location points for charging stations in the corresponding road network.

\subsection{Computing the reachability graph}
The following procedure is used to compute the reachability graphs $G^r_t$. First, the vertices of $G^s$ are copied into $G^r_t$. Next, for each vertex $v$ in $G^r_t$, we add an edge between $v$ and all the vertices in $G^r_t$ which are within the distance $t$ from $v$ in $G^s$. Here the distance between two vertices of $G^s$ is measured as the length of a shortest path. This is accomplished by performing a modification of the breadth-first search from the source vertex $v$ in $G^s$. Specifically, we employ Dijkstra's algorithm, but terminate the search when all vertices within a network distance $t$ of $v$ have been found. In our simulations, the graph $G^s$ is sparse. Therefore, to minimize running time, we implemented Dijkstra's algorithm using a binary heap based priority queue. This gives a running time of $O((|V^s|+|E^s|)\log|V^s|)$ for each call of this algorithm \cite{BH15}. This algorithm is called for each $v\in V^s$ as the source vertex, giving a total running time of $O(|V^s|(|V^s|+|E^s|)\log|V^s|)$ for computing the reachability graph.

\subsection{Computing $k$-dominating sets in reachability graphs}
Algorithm \ref{alg1} below is a randomized heuristic to compute a small-size $k$-dominating set in $G^r_t$ and is an adjustment of the corresponding randomized algorithm from \cite{GPZ2013}. It uses as an input a reachability graph $G^r_t$ and a positive integer $k$, $k\le \delta(G^r_t)$. Algorithm \ref{alg1} returns a (minimal by inclusion) $k$-dominating set $D$ in $G^r_t$, which provides a set of locations for charging stations in $G^s$ such that, from any given point (vertex) in $G^s$, a driver has at least $k$ different feasible options to reach a charging station when the remaining driving battery charge is enough for $t$ kilometers. In general, the order of the $k$-dominating set $D$ in $G^r_t$ returned by Algorithm \ref{alg1} satisfies the upper bound of Theorem \ref{k-dom} with a positive probability, i.e. the expectation of the order of $D$ satisfies the upper bound of Theorem \ref{k-dom}.

The upper bound of Theorem \ref{k-dom} is known to be asymptotically best possible for general graphs on $n$ vertices in the case of $1$-dominating sets (e.g., see \cite{A1990}). In general, it is currently one of the best bounds for $\gamma_k(G)$ and likely to be asymptotically best possible for arbitrary $k$, $1\le k\le \delta$. However, it turns out that the bound of Theorem \ref{k-dom} is not sharp enough in the case of particular reachability graphs of road networks for Boston and Dublin. As a result, randomized Algorithm \ref{alg1} usually returns a non-minimal $k$-dominating set of an unreasonably large size.
Therefore, instead of using the minimum vertex degree $\delta(G^r_t)$ of the reachability graphs $G^r_t$ to compute the probability $p$ and parameters $\delta'$ and $b_{k-1}$ in Algorithm \ref{alg1}, we use in our experiments the average vertex degree of $G^r_t$, i.e.
$$
\bar{d}(G^r_t)=\frac{1}{n}\sum_{i=1}^{n} d_i.
$$
In general, using $\bar{d}(G^r_t)$ instead of $\delta(G^r_t)$ in Algorithm \ref{alg1} doesn't guarantee obtaining a $k$-dominating set satisfying the upper bound of Theorem \ref{k-dom}. However, in the particular cases of road networks of Boston and Dublin, using $\bar{d}(G^r_t)$ in Algorithm \ref{alg1} provides good computational results satisfying the upper bound of Theorem \ref{k-dom} as well. Notice that we have $k\le \delta(G^r_t)\le \bar{d}(G^r_t)$.
Our implementation of randomized Algorithm \ref{alg1} is enhanced with some other heuristics as well, and we run it several times to obtain smaller size $k$-dominating sets in $G_t^r$.\\

\incmargin{-0.1em}
\restylealgo{ruled}
\begin{algorithm}\label{alg1}
    \caption{Randomized $k$-dominating set}

    \KwIn{A reachability graph $G^r_t$ and an integer $k$, $k\le \delta$.}
    \KwOut{A $k$-dominating set $D$ of $G^r_t$.}
    \BlankLine
\Begin{

Compute the probability 
$\displaystyle{p = 1-\frac{1}{\sqrt[\delta']{{b}_{k-1}{(1 + \delta')}}}}$\;
\SetLine
Initialize set $A=\emptyset$;\tcc*[f]{Form a set $A \subseteq V(G^r_t)$}\\
 \ForEach{vertex $v\in V(G^r_t)$} {
    with the probability $p\;$, decide whether $v\in A$, otherwise $v\not\in A$; \tcc*[h]{this forms a subset $A\subseteq V(G^r_t)$}\;
}
\SetLine
Initialize $B=\emptyset$;\\
 \ForEach{vertex $v\in V(G^r_t)\backslash A$} {
    \If{$|N(v)\cap A|<k$}{
        \tcc*[h]{$v$ is dominated by less than $k\;$ vertices of $A$}\\
        add $v$ into $B$; \tcc*[h]{this forms a subset $B\subseteq V(G^r_t)\backslash A$}\;
    }
}
Put $D = A \cup B$;\tcc*[f]{$D$ is a $k$-dominating set in $G^r_t$}\\
If possible, remove some vertices from $D$ to have a minimal $k$-dominating set  $D'$ in $G^r_t$;\\
\Return $D'$\; }
\end{algorithm}

In the experiments, we have compared the results obtained by using the randomized approach of Algorithm \ref{alg1} with those returned by a simple recursive greedy method described in Algorithm \ref{alg:greedy}.
Notice that, when $k=1$, Algorithm \ref{alg:greedy} is a simple deterministic (greedy) approach derandomizing Algorithm \ref{alg1} (e.g., see \cite{AS1992}). However, as already suggested by the results for $k=2$ in \cite{HH2008}, it can be a marvellous task to derandomize Algorithm \ref{alg1} or similar randomized algorithms in general. The results returned by Algorithm \ref{alg:greedy} have been used as a benchmark to run Algorithm \ref{alg1} several times to obtain better results (all satisfying the upper bound of Theorem \ref{k-dom}).\\

\begin{algorithm}\label{alg:greedy}
	\caption{Greedy $k$-dominating set}
	\KwIn{A reachability graph $G^r_t$, an integer $k$, and $D \subseteq V(G^r_t)$.}
	\KwOut{A $k$-dominating set $D$ of $G^r_t$.}
	\While{$ |\lbrace v\in V(G^r_t)\backslash D : |N(v)\cap D|<k \rbrace| > 0 $}{
		Set\,\  $U = \lbrace\, v\in V(G^r_t)\backslash D\ :\ |N(v)\cap D|<k\, \rbrace$;\\
		Find\,\  $\displaystyle u =\, \argmax_{v\in V(G^r_t)\backslash D}\, |N(v)\cap U|$; \\
		Put\,\ $D = D \cup \lbrace u \rbrace$;
	}
\end{algorithm}

The original $k$-dominating sets returned by Algorithms \ref{alg1} and \ref{alg:greedy} are normally not minimal (by inclusion). Therefore, we have used a simple greedy procedure to reduce them to minimal $k$-dominating sets and to check that the final $k$-dominating sets are minimal. A pseudocode for this elimination of redundancy is presented in Algorithm \ref{alg:minimal}.\\

\begin{algorithm}\label{alg:minimal}
	\caption{Minimal $k$-dominating set}
	\KwIn{A reachability graph $G^r_t$ and a $k$-dominating set $D$ of $G^r_t$.}
	\KwOut{A minimal $k$-dominating set $D$ of $G^r_t$.}
	Order the vertices in $D$ as $L = (v_1, \dots, v_{|D|})\ :\ v_i \in D,\ |N(v_i) \setminus D| \leq |N(v_{i+1}) \setminus D| $; \\
	\For{$i = 1$ to $n$}{
	\If{$D \setminus \lbrace v_i \rbrace$ is $k$-dominating set of $G^r_t$}{
		Put\,\ $D = D \setminus \lbrace v_i \rbrace$;
		}
	}
\end{algorithm}

\subsection{Complexity analysis for the randomized algorithm}
Computing the binomial coefficient $b_{k-1}$ in Algorithm \ref{alg1} is normally done by using the dynamic programming and Pascal's triangle, which has $O(\delta^2)$ time complexity in this case. The minimum vertex degree $\delta$ of $G^r_t$ can be computed in linear time in the number of edges $m$ of $G^r_t$. Notice that $O(\delta^2)$ does not exceed $O(m)$. Therefore, computing probability $p$ can be done in $O(m)$ time.
It takes $O(n)$ time to find the set $A$, where $n$ is the number of vertices in $G^r_t$.
The numbers $z=|N(v)\cap A|$ for each vertex $v\in V(G^r_t)\backslash A$ can be computed separately or when finding the set $A$. We need to keep track of them only while $z<k$. Since we may need to browse through all the neighbours of
vertices in $A$, in total, it can take $O(m)$ steps to calculate
all the necessary $z$'s for all $v\in V(G^r_t)\backslash A$. Then the set
$B$ can be also found in $O(n)$ steps. Thus, in total, Algorithm
\ref{alg1} runs in $O(m+n)$ time. Since we heuristically use the average vertex degree $\bar{d}$ instead of the minimum vertex degree $\delta$ in our experiments with Algorithm \ref{alg1}, the complexity analysis of its implementation is slightly different, but can be easily derived from the analysis above.

It is possible to use simple heuristics when computing set $B$ in Algorithm \ref{alg1}. First, we can build set $B$ recursively, considering the undercovered vertices in $G^r_t\backslash A$ one by one. Then, we may want to include the most undercovered vertices, i.e. vertices $v\in G^r_t\backslash A$ with the smallest intersection $|N(v)\cap A|$, into $B$ first, and update set $A$ gradually by including the new vertices from $B$ directly into $A$ to form $A'$. This would recursively update the numbers $z=|N(v)\cap A'|$, make some of the undercovered vertices covered enough with at least $k$ neighbours in $A'$, and increase the coverage score $z=|N(v)\cap A'|$ for some $v\in G^r_t\backslash A'$ (in comparison to $|N(v)\cap A|$) to influence selection of the next vertex for $B$ in iteration. A heuristic ``greedy extension" of $A$ procedure to find the sets $A'$ is similar to Algorithm \ref{alg:greedy}.

Finally, since the initial recursively obtained $k$-dominating set $A'$ may be not minimal, we try to exclude some vertices from $A'$. This is implemented by removing vertices from $A'$ one by one and checking whether the $k$-domination property still holds. A heuristic procedure to guarantee the minimality of the returned $k$-dominating set is described in Algorithm \ref{alg:minimal}.

\section{Experimental evaluation}
\label{sec:experimental_evaluation}

In this section,
in order to evaluate the proposed methodology, we describe our experiments with multiple domination models and corresponding algorithms in the case of two road network graphs $G^s$ corresponding to the cities of Boston in the USA and Dublin in Ireland. 

\subsection{Data}
\begin{figure}
\centering     
\subfigure[]{\label{fig:b}\includegraphics[width=6.5cm]{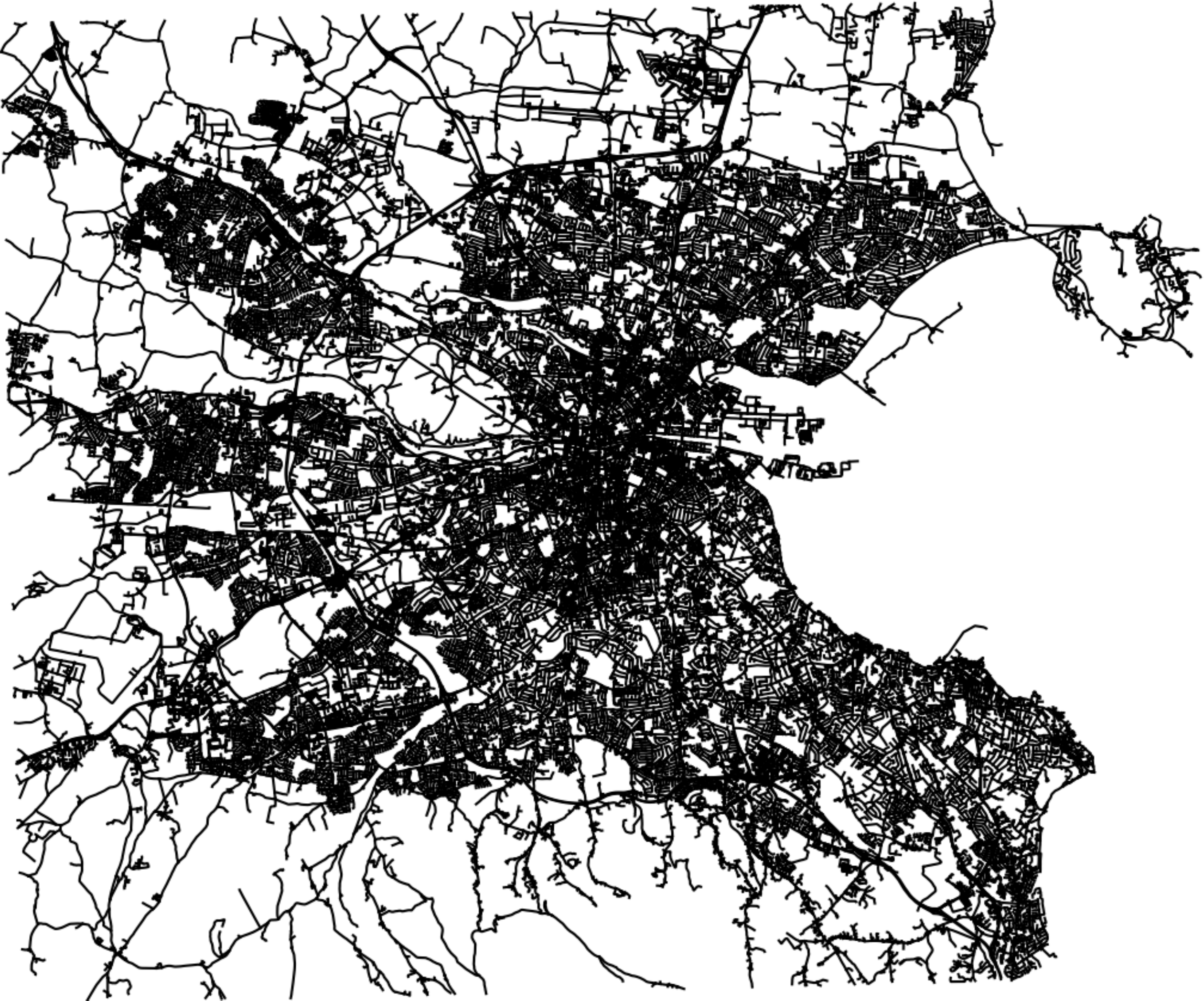}
\label{fig:dublin_network}}
\hspace{0mm}
\subfigure[]{\label{fig:a}\includegraphics[width=6.5cm]{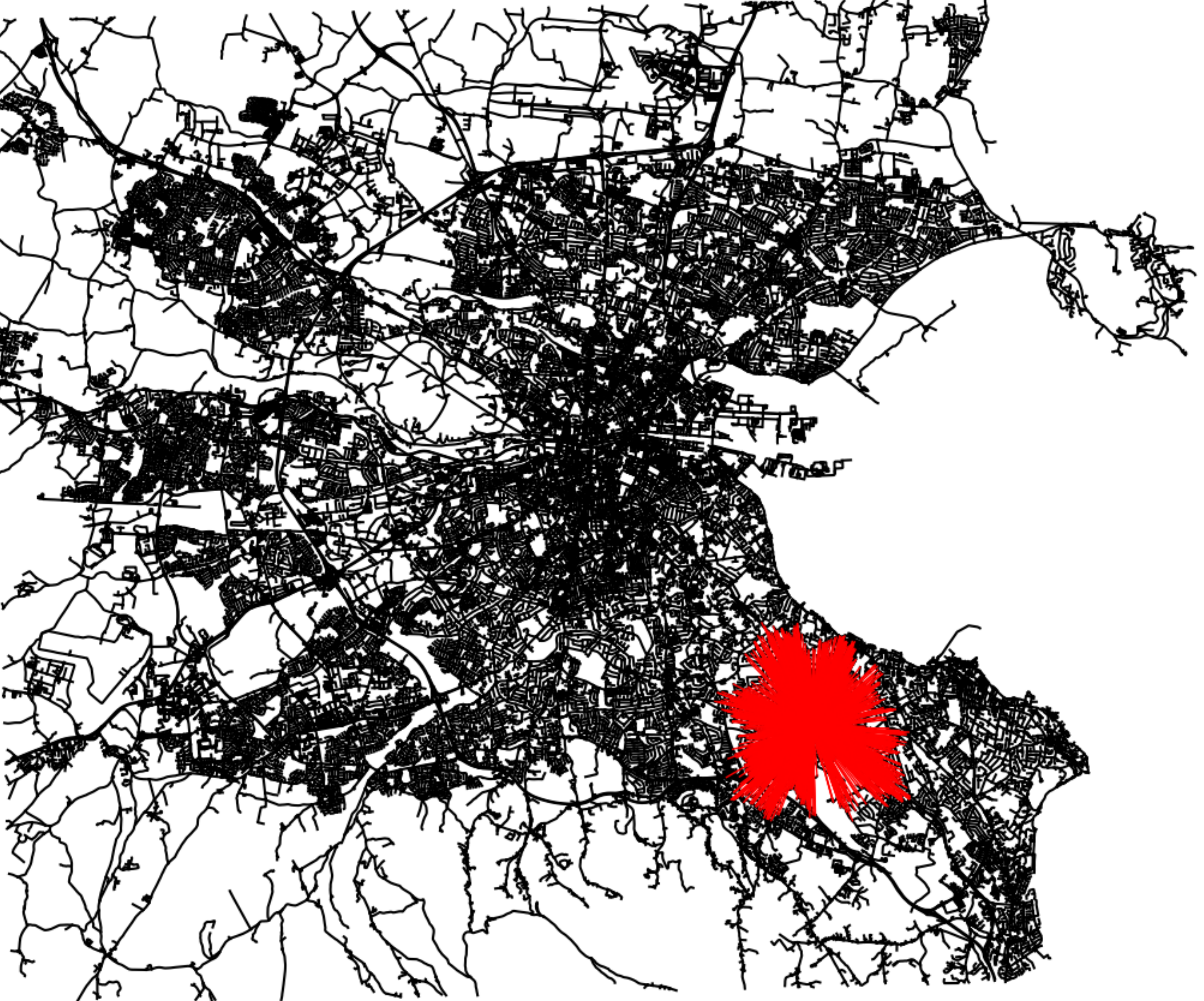}
\label{fig:Dublin_reach_3}}
\caption{(a)\,The road network for the city of Dublin; (b)\,Neighbourhood of a vertex in the corresponding reachability graph.}
\label{fig:Dublin_reachability_vertex}
\end{figure}

The two road networks in question are illustrated in Figures \ref{fig:Boston_street_network} and \ref{fig:dublin_network}, respectively, and are obtained from OpenStreetMap \cite{Cor2013}. The graph $G^s$ corresponding to Boston consists of $21,542$ vertices and $31,112$ edges. It is contained within a rectangular region of width $15.5$km and height $12.1$km. The graph $G^s$ corresponding to Dublin consists of $55,162$ vertices and $64,437$ edges. It is contained within a rectangular region of width $29.5$km and height $24.6$km. Notice that both road network graphs are either planar or ``almost" planar: when considering them embedded in the plane as road maps, the edge crossings are only possible in the case of road bridges and tunnels. Moreover, these two graphs are sparse in terms of the number of edges $m$, which satisfies the linear upper bound in terms of the number of vertices for planar graphs, $m\le 3n-6$, $n=|V(G^s)|$, as opposite to the general worst case quadratic upper bound $m\le\frac{n(n-1)}{2}$, i.e. $m=O(n^2)$.

The most appropriate reachability threshold $t$ for the reachability graph $G^r_t$ is a function of a large number of parameters. This includes the number of electrical vehicles which require charging, the number of charging stations one is able to install, the number of charging options one wishes to offer, and the cost of installing a charging station. Determining this threshold would probably best be done by consultation with city planners. In this paper, we assume the most appropriate reachability threshold for both cities' road networks and electrical vehicles is set to be $3.0$km.

For each road network graph $G^s$, we computed the corresponding reachability graph $G^r_{3.0}$. These graphs are illustrated in Figures \ref{fig:reachability_vertex} and \ref{fig:Dublin_reach_3}. The reachability graph $G^r_{3.0}$ corresponding to Boston contains $21,542$ vertices and $23,052,466$ edges (approx. $9.94\%$). The reachability graph $G^r_{3.0}$ corresponding to Dublin contains $55,162$ vertices and $54,306,700$ edges (approx. $3.57\%$). All $k$-dominating sets are computed using these two reachability graphs.

\subsection{Cardinality of $k$-dominating sets}
\begin{figure}
\centering     
\subfigure[]{\label{fig:b}\includegraphics[width=6.4cm]{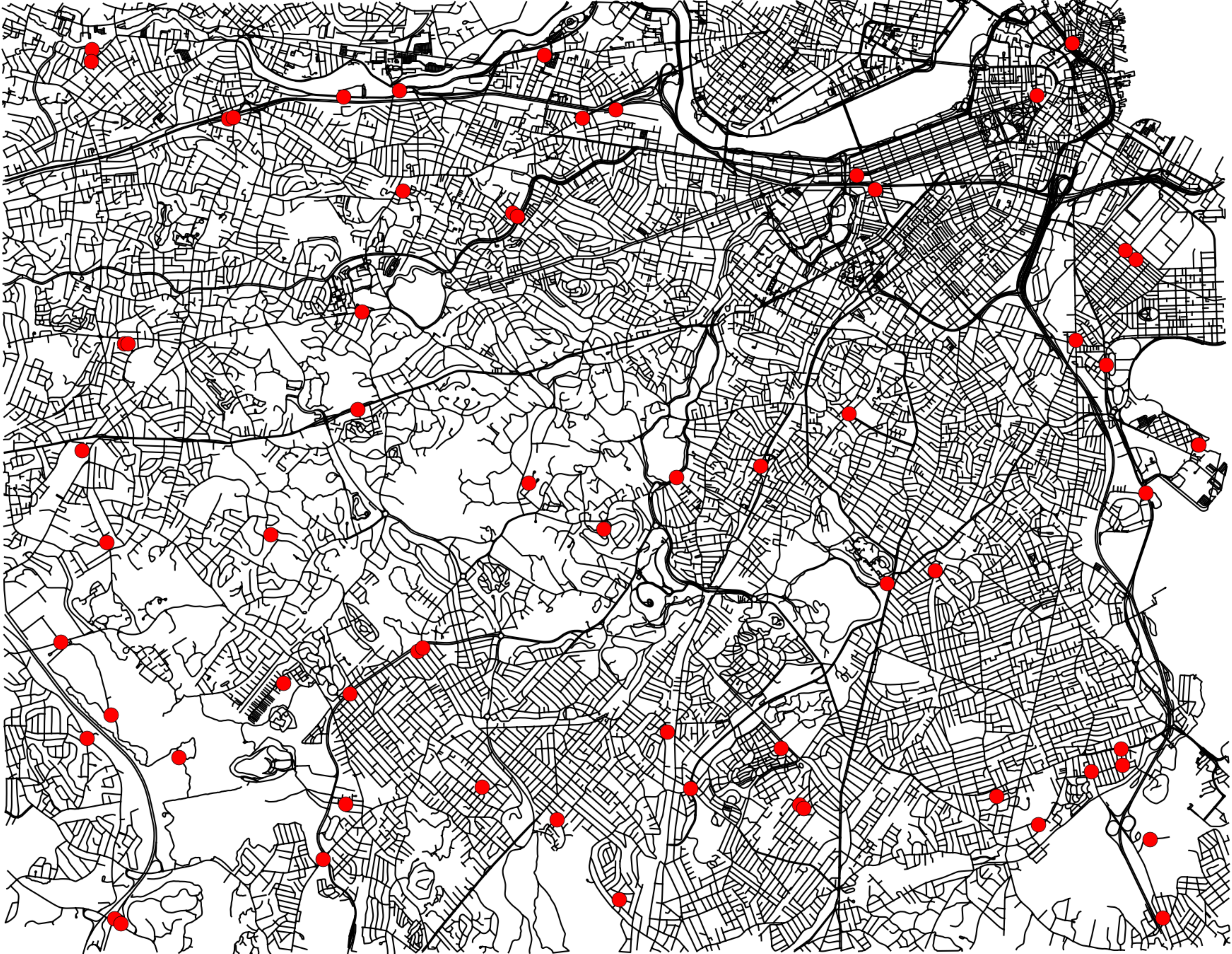}
\label{fig:dominating_set_Boston_k2_r30_greedy}}
\hspace{1mm}
\subfigure[]{\label{fig:a}\includegraphics[width=6.4cm]{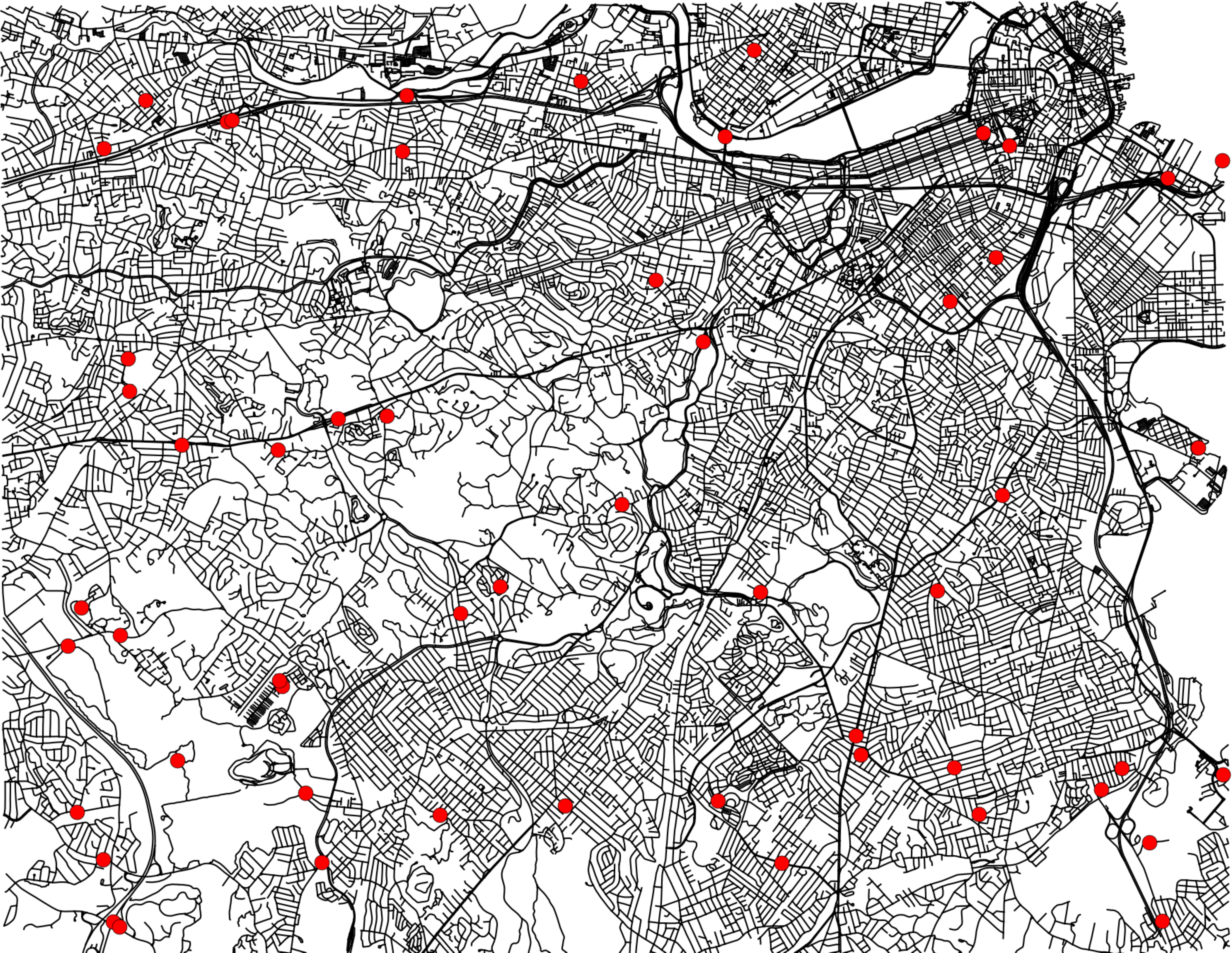}
\label{fig:dominating_set_Boston_k2_r30_random_9}}
\caption{(a) The $2$-dominating set of $64$ vertices computed by the greedy algorithm and (b) the smallest $2$-dominating set of $56$ vertices computed by the randomized algorithm (both for the city of Boston; the vertices in the $2$-dominating sets are in red; see Table \ref{tab:k_dominating_set_size}).}
\label{fig:Boston_dominating_set}
\end{figure}

\begin{figure}
\centering     
\subfigure[]{\label{fig:b}\includegraphics[width=6.4cm]{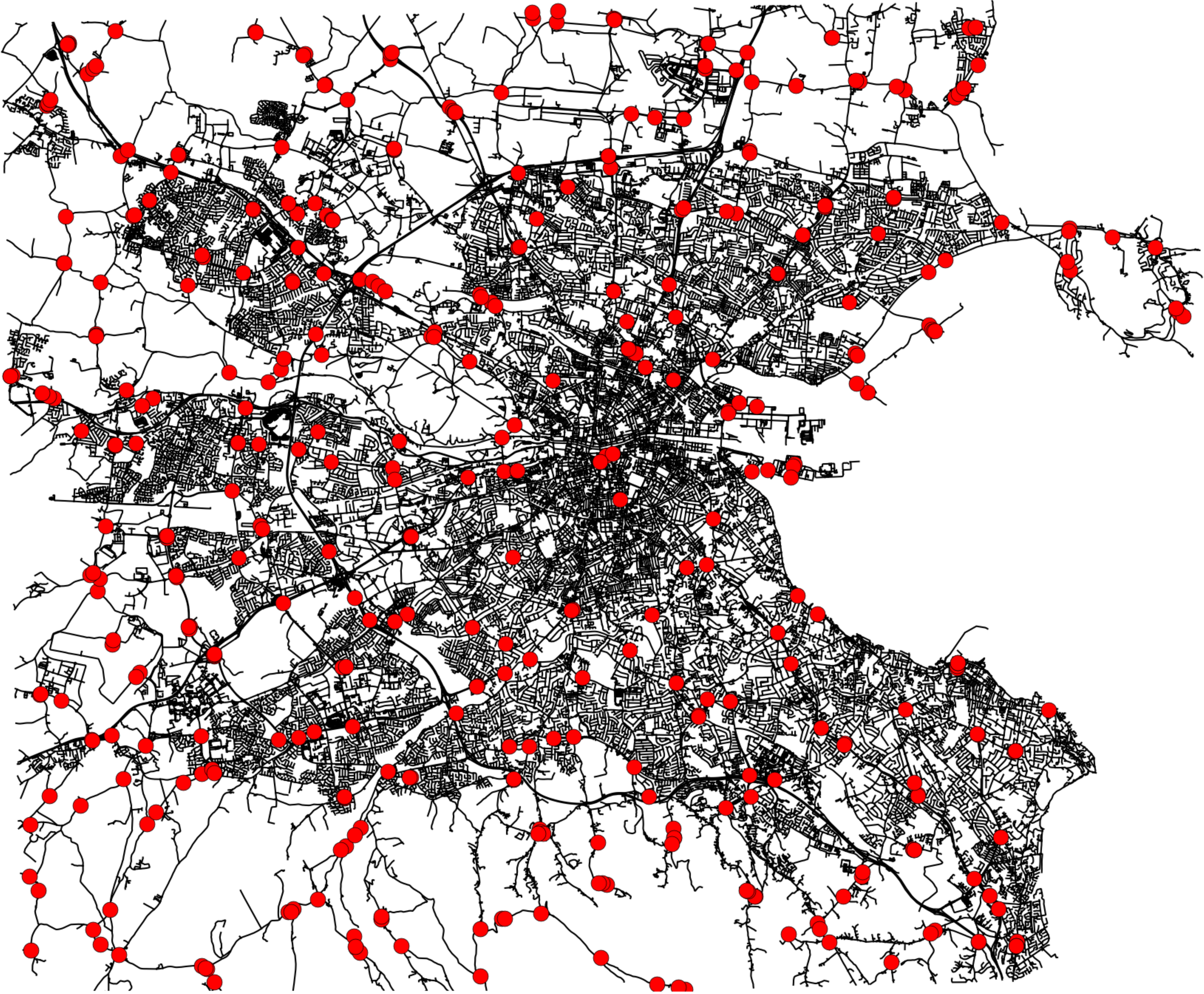}
\label{fig:dominating_set_Dublin_k4_r30_greedy}}
\hspace{1mm}
\subfigure[]{\label{fig:a}\includegraphics[width=6.4cm]{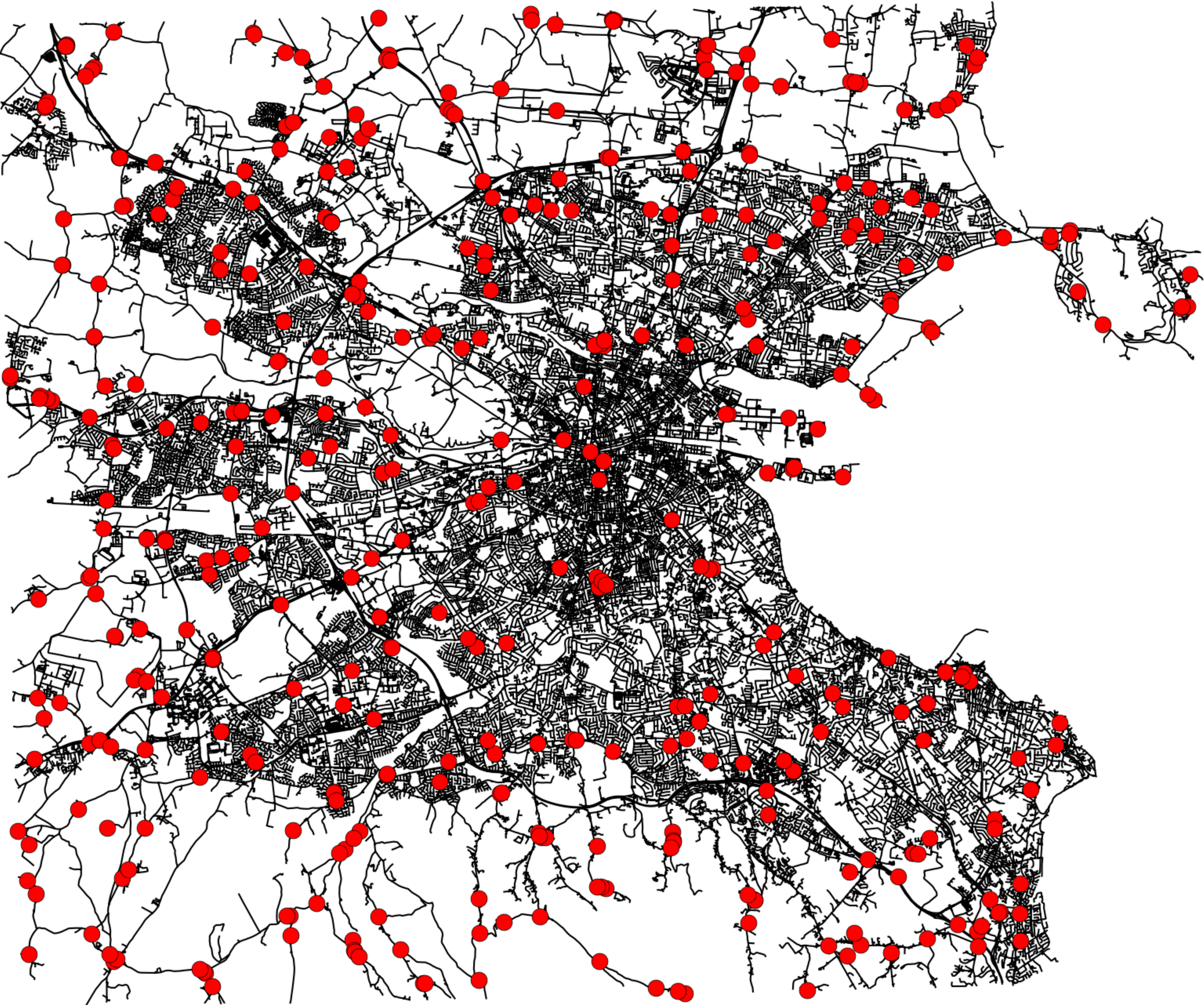}
\label{fig:dominating_set_Dublin_k4_r30_random_8}}
\caption{(a) The $4$-dominating set of $413$ vertices computed by the greedy algorithm and (b) the smallest $4$-dominating set of $411$ vertices computed by the randomized algorithm (both for the city of Dublin; red dots indicate the vertices in the $4$-dominating sets; see Table \ref{tab:k_dominating_set_size}).}
\label{fig:Dublin_dominating_set}
\end{figure}

For each of the reachability graphs $G^r_{3.0}$, one $k$-dominating set was computed using the greedy algorithm, and ten $k$-dominating sets were computed using the randomized algorithm for $k=1,2,4$. Table \ref{tab:k_dominating_set_size} displays the cardinalities of the $k$-dominating sets computed using the greedy algorithm and the cardinalities of the smallest $k$-dominating sets computed using the randomized algorithm for each of the cities and each value of $k=1,2,4$. In four out of the six cases, the randomized algorithm computed a smaller dominating set than the same multiplicity dominating set returned by the greedy algorithm. The two $2$-dominating sets for the city of Boston are displayed in Figure \ref{fig:Boston_dominating_set}. The two $4$-dominating sets for the city of Dublin are displayed in Figure \ref{fig:Dublin_dominating_set}. 

A visual inspection of Figures \ref{fig:Boston_dominating_set}, \ref{fig:Dublin_dominating_set}, and others reveals that spatial locations of the elements in the dominating sets tend to be more spatially clustered when computed using the greedy algorithm. This can be attributed to the greedy nature of the approach: vertices of high degree in the corresponding reachability graphs tend to be spatially clustered, and the greedy algorithm will add these high degree vertices to the dominating set first. On the other hand, the randomized algorithm initially adds a random set of vertices to the future dominating set, and these vertices are likely to be spatially distributed in a uniform way.

\begin{table}
\centering
\begin{tabular}{|c|c|c|c|c|}
 \hline 
Network & \multicolumn{2}{|c|}{Boston} & \multicolumn{2}{|c|}{Dublin} \\ 
 \hline 
Algorithm & {\hspace{2mm} Greedy \hspace{2mm}} & Randomized & \hspace{1mm} Greedy \hspace{1mm} & Randomized \\ 
 \hline 
 $k=1$ & 32 & 31 & 110 & 111 \\ 
 \hline 
 $k=2$ & 64 & 56 & 214 & 215 \\ 
 \hline 
 $k=4$ & 122 & 115 & 413 & 411 \\ 
 \hline 
 \end{tabular}
\caption{Cardinalities of the $k$-dominating sets computed using the greedy algorithm and the smallest $k$-dominating sets computed using the randomized algorithm for each city and each value of $k$.}
\label{tab:k_dominating_set_size}
\end{table} 

\subsection{Reachability of stations}
Given a fixed $k$-dominating set $X$ in a reachability graph $G^r_t$ corresponding to a road network graph $G^s$, the number of elements in $X$ reachable from a given vertex in $G^s$ is a non-decreasing function of distance. To examine this phenomenon, we consider the smallest $2$-dominating sets computed using the randomized algorithm for Boston and Dublin. These two sets contain $56$ and $215$ elements, respectively. The set corresponding to Boston is illustrated in Figure \ref{fig:dominating_set_Boston_k2_r30_random_9}. 

We computed the mean and standard deviation of the number of vertices in $X$ reachable from a vertex in $V(G^s)\backslash X$ as a function of distance. These values are displayed in Table \ref{tab:reachability_station}. An analysis of this table reveals the following facts. Despite the fact that the dominating sets were computed for a reachability graph with the reachability threshold of $3$km, the mean number of vertices in $X$ within a distance of $1$km of a vertex in $V(G^s)\backslash X$ for each of the cities is $0.5$. Furthermore, for both cities, the mean number of elements in $X$ within the distance of $3$km from a vertex in $V(G^s)\backslash X$ is significantly larger than $2$.
This means more support and flexibility than only $2$ a priory guaranteed options for recharging electrical vehicles in many points of these road networks.

On the other hand, increasing the reachability threshold to some $q>t=3$km and keeping the set of vertices $X$ fixed in $G^r_q$ should allow us to increase the minimum multiplicity of coverage of each vertex $v\in V(G^s)\backslash X$ by the vertices in $X$ to have $X$ as a $k$-dominating set with $k>2$ in $G^r_q$. We have computed the minimum multiplicity of coverage by the same $2$-dominating sets $X$ in the corresponding reachability graphs $G^r_q$ with the reachability threshold $q$ increasing to $4,\,5$, and $6$km. As shown in the corresponding columns of Table \ref{tab:reachability_station}, this increase of the reachability threshold haven't allowed us to increase the minimum number of options for the city of Boston, but have turned the $2$-dominating set $X$ of $G^r_{3.0}$ into a $3$-dominating set in $G^r_{6.0}$ for the city of Dublin. In other words, the drivers in Dublin are going to have at least $3$ options available within the distance of $6$km for recharging the batteries when using the same $2$-dominating set $X$ from $G^r_{3.0}$.

\begin{table}
\centering
\begin{tabular}{|c|c|c|c|c|c|c|}
 \hline 
Network & \multicolumn{3}{|c|}{\hspace{1mm} Boston} & \multicolumn{3}{|c|}{Dublin} \\ 
 \hline 
Stats & {\hspace{1mm} Mean \hspace{1mm}} & Std & Min & \hspace{1mm} Mean \hspace{1mm} & Std & Min \\ 
 \hline 
 1 km & 0.5 & 0.7 & 0 & 0.5 & 0.7 & 0 \\
 \hline 
 2 km & 1.9 & 1.1 & 0 & 2.0 & 1.1 & 0 \\
 \hline 
 3 km & 4.3 & 1.4 & 2 & 4.6 & 1.5 & 2 \\
 \hline
 4 km & 7.3 & 1.9 & 2 & 8.5 & 2.3 & 2 \\
 \hline
 5 km & 10.9 & 2.6 & 2 & 13.7 & 3.1 & 2 \\
 \hline
 6 km & 14.9 & 3.7 & 2 & 20.0 & 4.2 & 3 \\
 \hline
 \end{tabular}
\caption{Three statistics for the number of charging stations reachable from a vertex outside of the charging station locations for the road networks of Boston and Dublin computed as a function of distance.}
\label{tab:reachability_station}
\end{table}

\subsection{Detour required}
The number of options available for recharging electrical vehicles in a road network $G^s$ increases as a function of the multiplicity value $k$ in the corresponding $k$-dominating set. In turn, this may reduce the length of detours required for recharging electrical vehicles. To quantify this phenomenon, we consider the situation where a driver of an electrical vehicle wishes to travel from a source location to a destination, but first needs to have their vehicle recharged. Therefore, the driver considers all charging stations within the distance of $3$km from the source and charges their vehicle at a charging station which minimizes the detour. Here the detour is the difference between the distance from source to destination and the sum of distances from source to the charging station and from the charging station to destination. 

\begin{figure}
\begin{center}
\includegraphics[width=8cm]{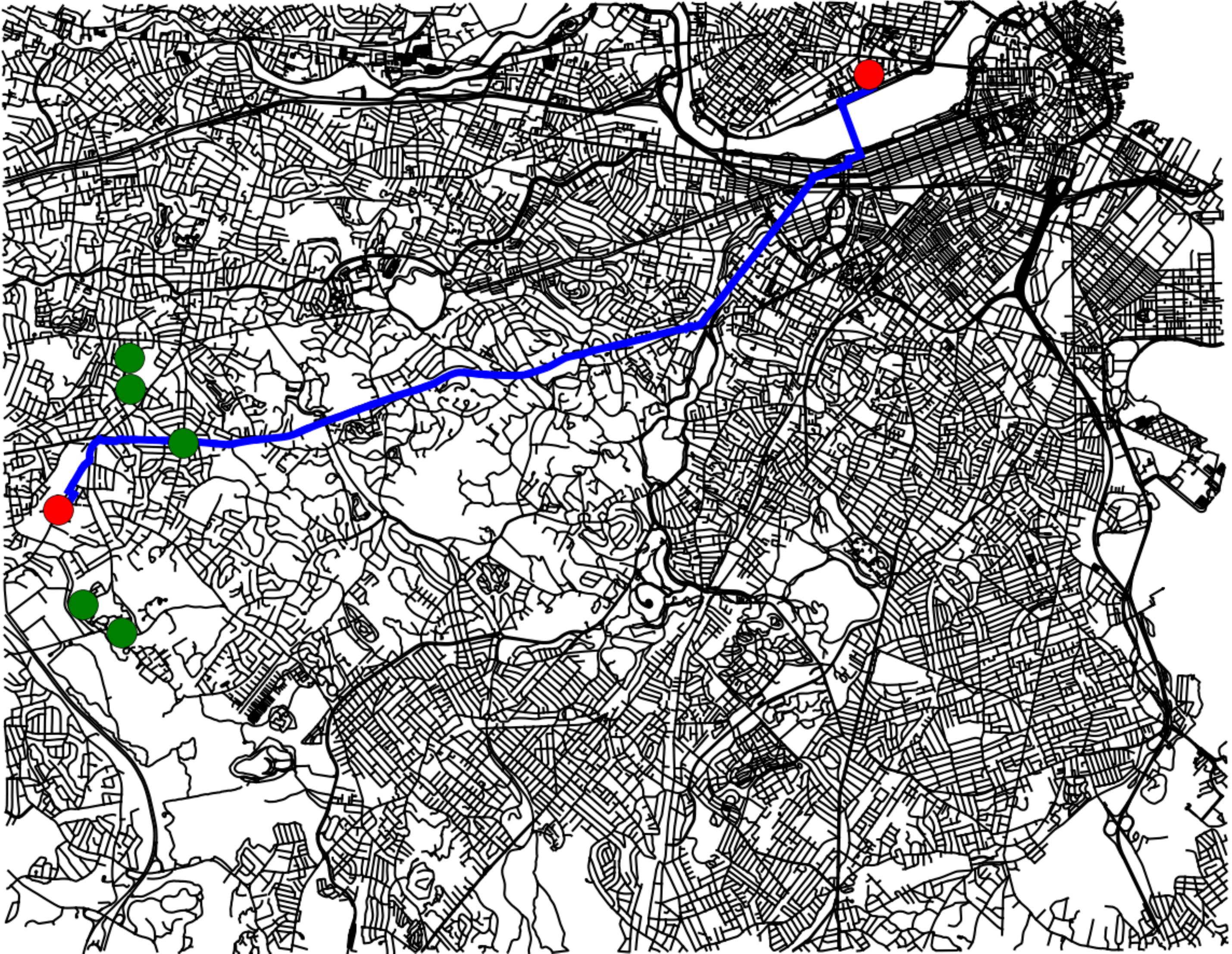}
\caption{A detour through a charging station for the city of Boston.} 
\label{fig:detour_sv16310_ev17245}
\end{center}
\end{figure}

To illustrate this, consider Figure \ref{fig:detour_sv16310_ev17245} and the situation where the source and destination are represented by red dots in the left and upper right of the figure, respectively. Considering the smallest $2$-dominating set computed using the randomized algorithm ($56$ vertices, see Table \ref{tab:k_dominating_set_size} and Figure \ref{fig:dominating_set_Boston_k2_r30_random_9}), there are five charging stations within the distance of $3$km from the source. These five charging stations are represented by green dots in Figure \ref{fig:detour_sv16310_ev17245}. The route which minimizes the detour is represented by the blue line in the figure, and the detour in question is only $92$ meters.

For each of the cities of Boston and Dublin, we have selected two hundred random pairs of source and destination locations, and for each pair of the locations, the corresponding detour for recharging was calculated. Considering the smallest $k$-dominating sets computed using the randomized algorithm for different values of $k=1,2,4$, the corresponding mean and standard deviation of detours required for recharging are displayed in Table \ref{tab:detour}. As expected, for both cities, the mean and standard deviation values decrease as the multiplicity of domination parameter $k$ increases. 

\begin{table}
\centering
\begin{tabular}{|c|c|c|c|c|}
 \hline 
Network & \multicolumn{2}{|c|}{\hspace{1mm} Boston} & \multicolumn{2}{|c|}{Dublin} \\ 
 \hline 
Stats & {\hspace{1mm} Mean \hspace{1mm}} & Std & \hspace{1mm} Mean \hspace{1mm} & Std \\ 
 \hline 
 $k = 1$ & 769 & 777 & 747 & 863 \\
 \hline 
 $k = 2$ & 436 & 541 & 501 & 578 \\
 \hline 
 $k = 4$ & 316 & 415 & 298 & 465 \\
 \hline
 \end{tabular}
\caption{Statistics of detours required for recharging batteries for random pairs of source and destination locations in the road networks and different values of $k$.}
\label{tab:detour}
\end{table} 

\section{Conclusions}
\label{sec:conclusions}

In this paper, we show analysis and good suitability of multiple domination models for decision problems related to efficient and effective placement of charging stations for electrical vehicles in the road networks. These results can serve as a first approximation to more complicated mathematical models with real road networks and their constraints. We plan to develop this research in the direction of more subtle road and transportation network models, for example, using digraphs and $\alpha$-domination models.

The experimental results with the road networks of Dublin and Boston indicate that sensitivity of the upper bound of Theorem \ref{k-dom}, which is strong in general graphs, can be improved in particular cases. Therefore, we conjecture that more sensitive upper bounds similar to Theorem \ref{k-dom} can be obtained by considering the degree sequence of a graph and some other of its parameters and properties. In particular, it would be interesting to obtain a stronger version of Theorem \ref{k-dom} in the case of reachability graphs corresponding to planar or ``almost" planar graphs derived from the spacial layouts of road networks.\\





\noindent\textbf{References}

\begin{thebibliography}{00}


\bibitem{AS1992} N. Alon, J.H. Spencer,
The Probabilistic Method, 
John Wiley \& Sons Inc., New York, 1992.

\bibitem{A1990}
N. Alon, 
Transversal numbers of uniform hypergraphs, 
Graphs Combin. \textbf{6} (1990), 1--4.

\bibitem{BH15} H. Bast, D. Delling, A. Goldberg, M. M{\"u}ller-Hannemann, T. Pajor, P. Sanders, D. Wagner, R. Werneck,
Route planning in transportation networks,
Algorithm Engineering: Selected Results and Surveys, Lecture Notes in Computer Science \textbf{9220}, (2016), 19--80.

\bibitem{Cor2013} 
P. Corcoran, P. Mooney, M. Bertolotto,
Analysing the growth of OpenStreetMap networks,
Spatial Statistics \textbf{3} (2013), 21--32.

\bibitem{DHLM2000} J.E. Dunbar, D.G. Hoffman, R.C. Laskar, L.R. Markus,
$\alpha$-Domination,
Discrete Math. \textbf{211} (2000), 11--26.

\bibitem{FNS2015} S. Funke, A. Nusser, S. Storandt,
Placement of Loading Stations for Electric Vehicles: No Detours Necessary!,
Journal of Artificial Intelligence Research \textbf{53} (2015), 633--658.

\bibitem{GPZ2009} 
A. Gagarin, A. Poghosyan, V.E. Zverovich,
Upper bounds for $\alpha$-domination parameters,
Graphs Combin. \textbf{25} (2009), 513--520.

\bibitem{GPZ2013} 
A. Gagarin, A. Poghosyan, V. Zverovich,
Randomized algorithms and upper bounds for multiple domination in graphs and networks,
Discrete Appl. Math. \textbf{161} (2013), 604--611. 

\bibitem{HH2008} J. Harant, M.A. Henning,
A realization algorithm for double domination in graphs,
Util. Math. \textbf{76} (2008), 11--24.

\bibitem{LLC2014} A.Y.S. Lam, Y.-W. Leung, X. Chu,
Electric vehicle charging station placement: formulation, complexity, and solutions,
IEEE Transactions on Smart Grid \textbf{5} (2014), 2846--2856.

\bibitem{LC2013} J.K. Lan, G.J. Chang,
Algorithmic aspects of the $k$-domination problem in graphs,
Discrete Appl. Math. \textbf{161} (2013), 1513--1520.

\bibitem{PGHL2015} A. Poghosyan, D.V. Greetham, S. Haben, T. Lee,
Long term individual load forecast under different electrical vehicles uptake scenarios,
Applied Energy \textbf{157} (2015), 699--709.

\bibitem{SF2013} S. Storandt, S. Funke,
Cruising with a Battery-Powered Vehicle and Not Getting Stranded,
In Proc. of the $26th$ AAAI Conference on Artificial Intelligence, Vol. 3. (2012), 1628--1634.\\
 
\end{thebibliography}
\end{document}